 \documentclass[final,authoryear,3p,times,twocolumn]{elsarticle}
\usepackage{natbib}
\usepackage{graphicx}
\usepackage{epsfig,subfigure}
\usepackage{amssymb}
\usepackage{amsthm}
\usepackage{amsmath}
\def\astrobj#1{#1}
\journal{New Astronomy}
\begin{document}
\begin{frontmatter}
\title{BVRI Photometric and Polarimetric studies of W UMa type \\
Eclipsing Binary \astrobj{FO Hydra}}
\author[ddu]{Vinod Prasad\corref{cor1}}
\ead{vinod.pdr83@gmail.com}
\author[ari]{J. C. Pandey}
\ead{jeewan@aries.res.in}
\author[ddu]{Manoj K. Patel}
\ead{patelmanoj79@gmail.com}
\author[ddu]{D. C. Srivastava}
\ead{dcs.gkp@gmail.com}
\address[ddu]{Department of Physics, DDU Gorakhpur University, Gorakhpur-273009, India}
\address[ari]{Aryabhatta Research Institute of observational sciencES (ARIES), Manora Peak, Nainital, India 263 129}

\begin{abstract}
We present analysis of optical photometric and polarimetric observations of contact binary system \astrobj{FO Hydra} (\astrobj{FO Hya}). The computed period of the system is 0.469556$\pm$0.000003 days. An O-C curve analysis indicates an increase in its period by $5.77\times 10^{-8}$ day yr$^{-1}$. The photometric light curves are analyzed using Wilson-Devinney code (WD). The present analysis shows that \astrobj{FO Hya} is a B-subtype of W UMa type contact binary. The radii and mass of primary and secondary components are found, respectively, to be R$_{1}$ = 1.62$\pm$0.03 R$_{\odot}$ and R$_{2}$ = 0.91$\pm$0.02 R$_{\odot}$, and M$_{1}$ = 1.31$\pm$0.07 M$_{\odot}$ and M$_{2}$ = 0.31$\pm$0.11 M$_{\odot}$. The light curve shape shows small asymmetries around the primary and secondary maxima. This may be due to the presence of dark spots on the components. The polarimetric observations yield average values of its polarization to be 0.18$\pm$0.03, 0.15$\pm$0.03, 0.17$\pm$0.02 and 0.15$\pm$0.02 per cent in B, V, R and I bands, respectively. These values are appreciably lower than the typical polarization of \astrobj{W UMa} type binaries. We have discussed the possible sources of the observed polarization but in order to arrive at definitive answer to this more phase locked observations are needed.

\end{abstract}


\end{frontmatter}


\section{Introduction}

Eclipsing binary stars of W Ursae Majoris (W UMa)-type variables; referred to, in short, EW, have their light curves with strongly curved maxima and minima that are nearly equal in height/ depth. These stars are divided into two sub-classes: A-type and W-type. In the A-type systems the larger component has the higher temperature whereas in the W-type systems the smaller component has the higher temperature. However, the temperature difference in the primary and the secondary is less than 1000K. Observationally, it has been found that the A-type systems tend to have low mass ratios (q $<$ 0.3) and spectral type from A to F. W-type systems usually have mass ratios q $>$ 0.3 and spectral types of G or K. Besides, there is also another subclass of contact binaries having temperature difference of 1000 K or larger between their components (Rucinski, 1986; Eggleton, 1996). This temperature difference is considerably larger than the theoretical limit of about 1000 K (e.g. Ka{\l}u{\.z}ny, 1986; Hilditch and King, 1988). Lucy and Wilson (1979) introduced the class of B-type systems which are systems in geometrical contact, but not in thermal contact and therefore there are large surface temperature differences between the components. The B-type systems are sometimes referred to as poor thermal contact systems (Rucinski and Duerbeck, 1997). Further, the light curves of W UMa binaries usually show differences in brightness of their maxima; this asymmetry is called the O'Connell effect (O'Connell, 1951). The O'Connell effect has been of particular interest in understanding the light variations of W UMa binaries. It also has been reported that the shape of the light curves and the depth of eclipses vary with time for some systems. Thus, the asymmetric light curves of W UMa type systems   indicate the possible presence of starspots on one or both stars of the system. \\

The eclipsing binary \astrobj{FO Hydra} (BD-18 2822) has been discovered to be a variable by Hoffmeister (1936). First quantitative observations were made by Tsessevich (1954) who reported a period of 1.159 days. The later studies of \astrobj{FO Hya} indicate a slightly asymmetric light curve having strong curvature in both maxima and minima of unequal depths (Binnendijk, 1960; Kopal, 1978; Candy and Candy, 1997), which was explained by the presence of spot on a star of the binary system. Candy and Candy (1997) have suggested that the slight and irregular variation in the orbital period of \astrobj{FO Hya} is presumably caused by mass transfer between the components. They determine the orbital period to be 0.4695571 days. Recently, the light and radial velocity curve analysis was performed by Siwak et al. (2010). They find the contact nature of \astrobj{FO Hya} with temperature difference of $\sim$ 2333 K between components. However, their light curves have been incomplete at the first quadrature and because of this, they were unable to get O'Connel effect and other parameters exactly. Therefore, we have undertaken an investigation of this system in order to get a better parameters estimation. \\

In this paper, we report broad band B, V, R and I light curves analysis of the star \astrobj{FO Hya} using our photometric observations. We describe the observations and data reduction procedures in \S 2. In \S 3 we present the period analysis. The light curves and estimation of basic parameters are subject matter of \S 4 and \S 5 . In \S 6, we present the analysis of the observations using Wilson-Devinney light curve modeling technique (WD code). We discuss the sources of polarization in \S 7. The summary of the results is given in the last section. 

\section{Observations and Data Reduction}

\subsection{Optical Photometry}

The photometric observations during 11 nights between April 09, 2009 to March 04, 2010 have been carried out using broad band B, V, R and I filters  at Aryabhatta Research Institute of observational sciencES (ARIES), Nainital with a 2k$\times$2k CCD camera mounted on cassegrain focus of the 104-cm Sampurnanand telescope (ST). The exposure times are ranging from 10 to 90 seconds depending upon the individual system and the observing conditions. The CCD system consists of 24$\times$24 $\mu$$^{2}$ size pixel. The gain and readout noise of CCD are 10 e-/ADU and 5.3 e-, respectively. The CCD covers a field of view $\sim$13$\times$13 arcminute$^{2}$. Several bias and twilight flat frames were also taken during each observing run. Bias subtraction, flat fielding and aperture photometry were performed using IRAF\footnote{IRAF is distributed by National Optical Astronomy Observatories, USA; http://iraf.net}.\\
We have chosen 2mass 09594720-1908422 and 2mass 09594342-1905435, as the comparison and check stars, respectively. The image of the field depicting the  comparison and the check stars with \astrobj{FO Hya} at the center; taken from ESO online Digitized Sky Survey, is shown in Figure 1. The co-ordinates of the target, comparison and the check stars are listed in Table~\ref{table1}. Differential photometry, in the sense of variable minus comparison star, have been performed and the comparison-check light curve in V band along with the folded V-band light curve of \astrobj{FO Hya} is shown in Fig. 2. The ephemeris used is as given by equation (1) of the \S 3. From the light curve, it is clear that the comparison and check stars are constant thought the observations. The nightly mean of standard deviation ($\sigma$) of different measures of comparison minus check stars in B, V, R and I band are 0.019, 0.013, 0.012 and 0.011, respectively. 

\begin{figure}[h]
\begin{center}
{\includegraphics[width=7.0cm,angle=-360]{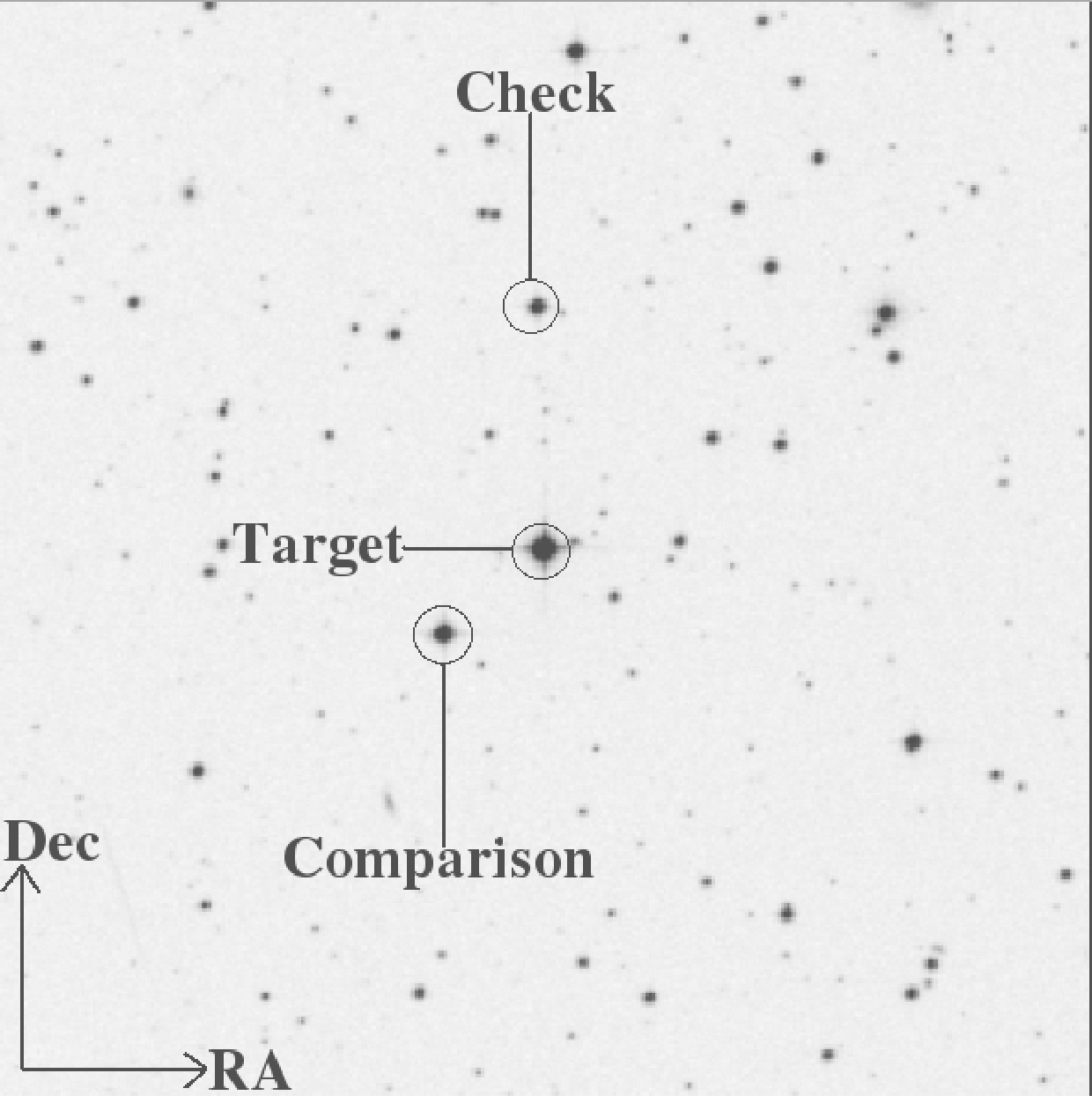}}
\vspace{-30mm}
\caption {A $\sim$ $10^{\prime}$ $\times 10^{\prime}$ image of the field with \astrobj{FO Hya} at center is taken from ESO online Digitized Sky Survey. The target, comparison and the check stars are marked.}
\label{id}
\end{center}
\end{figure}

\begin{table*}
\begin{center}
\caption{Coordinates and magnitudes of the program (\astrobj{FO Hya}), comparison (2Mass 09594720-1908422) and check (2Mass 09594342-1905435) stars.}
\label{table1}
\begin{tabular}{lccccc}
\hline 
Star & RA (J2000) & Dec (J2000) & B mag & V mag & J mag \\ \hline
\astrobj{FO Hya} & 09 59 43.29 & -19 07 56.22 & 11.31 & 11.05 & 10.36 \\
2Mass 09594720-1908422 & 09 59 47.21 & -19 08 42.27 & ... & ... & 11.41 \\
2Mass 09594342-1905435 & 09 59 43.43 & -19 05 43.54 & ... & ... & 11.99 \\
\hline 
\end{tabular} 
\end{center}
\end{table*} 

\subsection{Optical Polarimetry}

The broad-band B, V, R and I polarimetric observations of \astrobj{FO Hya} have been obtained on April 07, 23, 2010 and May 20, 2010 using ARIES Imaging Polarimeter (AIMPOL) mounted on the cassegrain focus of the 104-cm Sampurnanand Telescope of ARIES, Nainital. The detail descriptions about the AIMPOL, data reduction and calculations of polarization, position angle are given in Rautela et al. (2004). For calibration of polarization angle zero-point, we observed highly polarized standard stars, and the results are given in Table \ref{table2}. Both the program and the standard stars were observed during the same night. The obtained values of polarization and position angles for standard polarized stars are in good agreement with Schimidt et al. (1992). To estimate the value of instrumental polarization, unpolarized standard stars were observed. These measurements show that the instrumental polarization is below 0.1 per cent in all pass bands (see Table \ref{table2}). In fact, the instrumental polarization of ST has been monitored since 2004 within other projects as well (e.g. Medhi et al., 2007, 2008; Pandey et al., 2009). These measurements demonstrated that it is invariable in all B, V and R pass bands. The instrumental polarization was then applied to all measurements.

\begin{table}[ht]
\centering
\scriptsize
\caption{Observed polarized and unpolarized standard stars}
\label{table2}
\begin{tabular}{ccccccc}
\hline
\multicolumn{5}{c}{Polarized Standard}\\ 
\hline
& \multicolumn{2}{c}{Schmidt et. al (1992)}&\multicolumn{2}{c}{Present work} \\ 
\hline
Filter&P (per cent) & $\theta (^\circ)$ & P (per cent) & $\theta (^\circ)$ \\ 
\hline
\multicolumn{5}{c}{\underline{HD 154445}} \\                              
B & $3.445\pm0.047$ & $88.88\pm 0.39 $  & $3.836\pm 0.211$&$ 88.88  \pm 1.54 $ \\ 
V & $3.780\pm0.062$ & $88.79\pm 0.47 $  & $3.708\pm 0.098$&$ 88.79  \pm 0.78 $ \\ 
R & $3.683\pm0.072$ & $88.91\pm 0.56 $  & $3.652\pm 0.029$&$ 88.91  \pm 0.22 $ \\ 
I & $3.246\pm0.078$ & $89.91\pm 0.69 $  & $3.353\pm 0.006$&$ 89.91  \pm 0.05 $ \\ 
\multicolumn{5}{c}{\underline{HD 155197}}              \\      
B & $4.112\pm0.047$ & $103.06\pm 0.33$  & $4.109\pm 0.039$&$ 103.06 \pm 0.27$ \\ 
V & $4.320\pm0.023$ & $102.84\pm 0.15$  & $4.283\pm 0.006$&$ 102.84 \pm 0.04$ \\ 
R & $4.274\pm0.027$ & $102.88\pm 0.18$  & $4.285\pm 0.037$&$ 102.88 \pm 0.24$ \\ 
I & $3.906\pm0.041$ & $103.18\pm 0.30$  & $3.679\pm 0.060$&$ 103.18 \pm 0.45$ \\ 
\multicolumn{5}{c}{\underline{HD 154445}}              \\
B & $3.445\pm0.047$ & $88.88\pm 0.39 $  & $3.640\pm 0.084$&$ 88.88  \pm 0.65 $ \\
\multicolumn{5}{c}{\underline{HD 161056}}              \\
B & $3.799\pm0.055$ & $66.56\pm 0.42$  & $3.751\pm 0.055$&$ 66.56 \pm 0.41$ \\
\hline
& \multicolumn{2}{c}{Unpolarized Standard (\underline{$\beta$\ Uma)}} \\
& q(per cent) & u(per cent) \\
\hline
B & -\ 0.114 & -\ 0.091 \\
V & -\ 0.264 & -\ 0.133 \\
R & -\ 0.001 & -\ 0.070 \\ 
I & \ 0.005 & -\ 0.079 \\
\hline
\end{tabular}
\end{table}
\section{Period analysis}
\label{period}
There are number of techniques available for studying the periodicity in a time series data. We have derived the period of \astrobj{FO Hya} using CLEAN algorithm (Roberts et al., 1987) method. The power spectrum is shown in Fig. 3 wherein the peak frequency exits at $\sim$ 4.2592 (day$^{-1}$). Eclipsing binary light curves can be represented by two sine waves and therefore multiplying the period (=1/frerquency) so obtained by 2 gives appropriate orbital period of the binary system. We determine the orbital period of \astrobj{FO Hya} to be 0.46957$\pm$0.00004 day, which is in agreement with the period derived by Candy and Candy (1997) and Siwak et al. (2010).

\begin{figure}[h]
\begin{center}
{\includegraphics[width=7.5cm,angle=0]{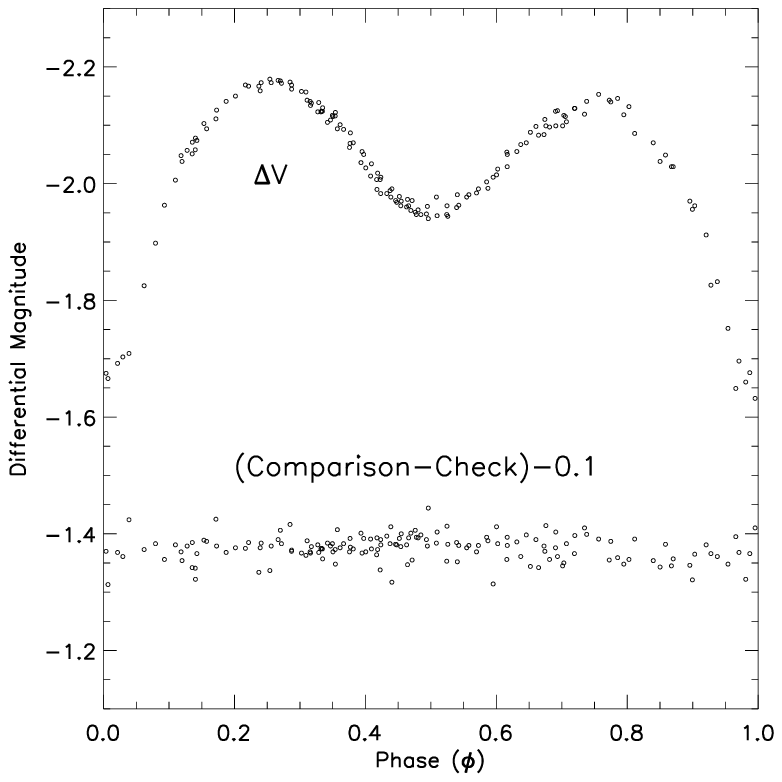}}
\caption {Folded V-band light curve of \astrobj{FO Hya} along with the Comparison-check folded light curve in V band. The ephemeris is given in equation 1.}
\label{C1C2}
\end{center}
\end{figure}

The timings of light minima are derived using Kwee and van Woerden's (1956) method. We derive the timings of the 10 minima, using our  quasi-simultaneous B, V , R and I bands observations and of the 25 minima, using All Sky Automated Survey (ASAS; Pojmanski, 2002) data. We also use the world-wide database of minima timings of Paschke and Brat (2006)\footnote{http://var.astro.cz/ocgate/} and obtain a total of 67 minima timings over a span of 65 years. The results are given in Table \ref{table3}.

\begin{figure}[h]
\begin{center}
{\includegraphics[height=8.0cm,angle=-90]{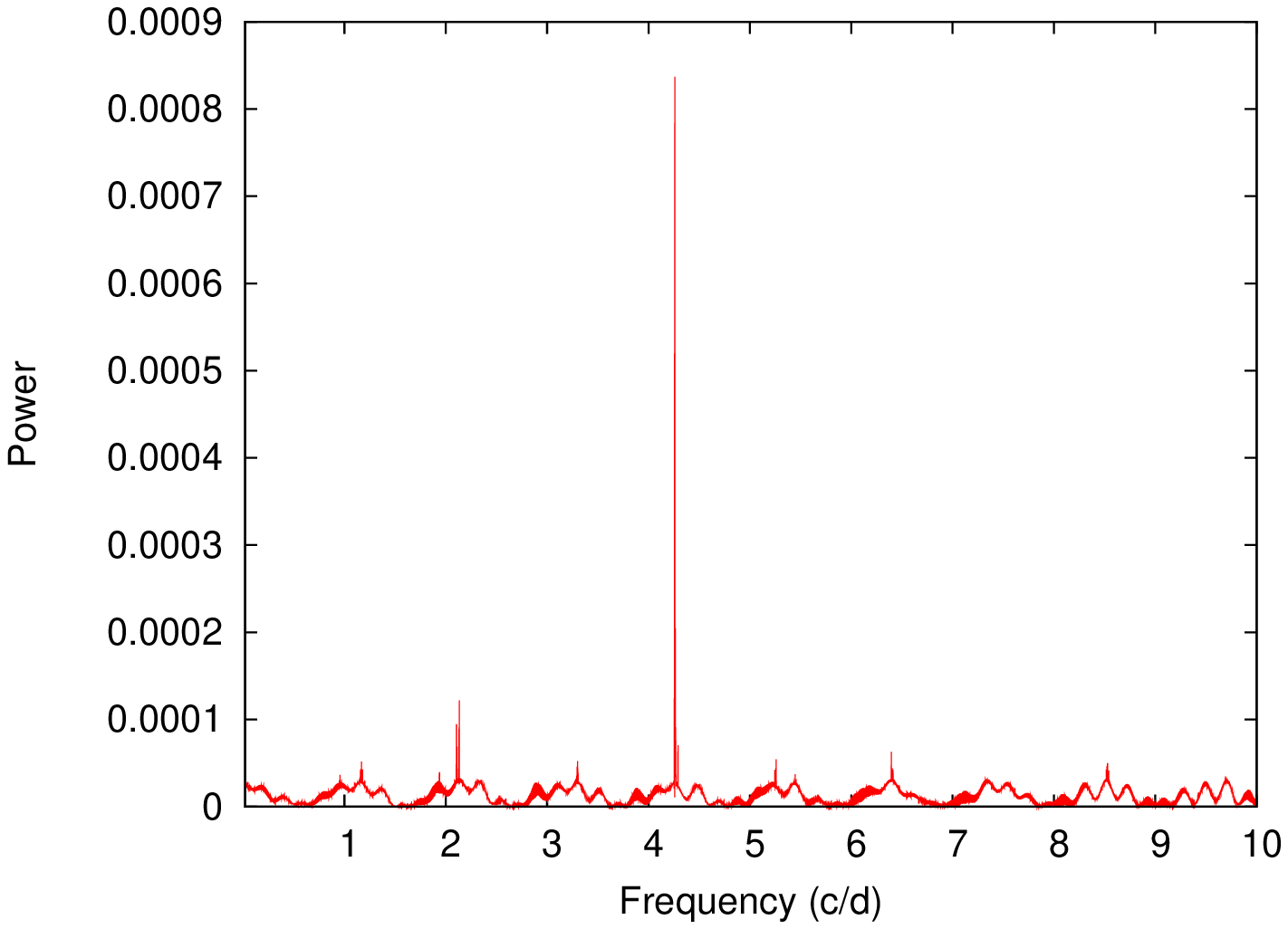}}
\caption {Power spectra of \astrobj{FO Hya} obtained from CLEAN algorithm. The most dominant peak in the spectra occurs to $\sim$4.259 c/d.}   
\label{pcl}
\end{center}
\end{figure}

We determine the ephemeris as given by the following equation 
\begin{equation}
Min.I(HJD) = 2455259.32959(64)+0^{d}{.469556}(3) \times E
\label{phmin}
\end{equation}

\noindent
The numbers given in parentheses represent the probable errors and are expressed in terms of the last quoted digits. Now we compute the O-C residuals, and plot its variation against the epoch number in Fig. 4 . As seen from the figure the general trend of the (O-C)$_{1}$ vs epoch diagram indicates a roughly parabolic distribution indicating a long time change in the orbital period. We, therefore, attempt second order polynomial to fit the data and represent it by a dashed line in the top panel of the Fig. 4. Here we also represent the residual from polynomial fit (O-C)$_{2}$ in the middle panel of the Fig. 4. It is noticed that there is still another change, which seemingly may be sinusoidal. The continuous curve in the middle panel of the Fig. 4 shows the best sinusoidal fit. We also plot the residuals of the sinusoidal fit, shown in the bottom panel of the figure, indicating no significant variation. The combined quadratic and sinusoidal fit for the observed data is obtained as:

\begin{equation}
\begin{split}
O-C = 0.0045(\pm0.0013)\\ 
+ (1.01\pm0.18)\times 10^{-6} \times E\\
+ (3.71\pm0.39)\times 10^{-11}\times E^{2}\\ 
+ 0.0041(\pm0.0018) sin[(1.91\pm0.18)\times 10^{-4} \times E\\
 + 3.93(\pm0.31)].\\
\end{split}
\label{qsine}
\end{equation} 

\noindent
The coefficient of the square term yields the rate of change of the period. We determine (dP/dt) to be 5.77 $\times 10^{-8}$ day yr$^{-1} $, which is equivalent to 7.42 $\times 10^{-11}$ day cycle$^{-1}$; meaning there by a period increase of 0.50 s century$^{-1}$ (see Qian and Ma (2001) for the details of the calculation). The change in the period is not unusual among W UMa type stars. The observed period increase/decrease rates of these W Uma type binaries are in the range from $\pm$10$^{-9}$ to $\pm$ 10$^{-6}$ day yr$^{-1}$ (Dryomova and Svechnikov, 2006). Several W UMa systems show the increasing trend in their orbital period e.g. \astrobj{YY Eri} (Kim et al., 1997), \astrobj{V839 Oph} (Akalin and Derman, 1997), \astrobj{AB And} (Kalimeris et al., 1994), \astrobj{DK Cyg} (Awadalla, 1994), \astrobj{V401 Cyg} (Herczeg, 1993), \astrobj{V566 Oph} (Maddox and Bookmyer, 1981). Examples of W UMa systems showing decreasing trend in their orbital period are \astrobj{TZ Boo}, \astrobj{Y Sex} (Qian and Liu, 2000), \astrobj{V502 Oph} (Derman and Demircan, 1992), \astrobj{U Peg} (Zhai et al., 1984), \astrobj{VW Cep} (Kaszas et al., 1998) and \astrobj{V781 Tau} (Liu and Yang, 2000a). \\ 

As the two components of \astrobj{FO Hya} are in a state of over-contact, we take the most likely cause of the period change as the transfer of mass between the components. The amount of mass transfer, dm is related to the change in period, dP, for a system of total mass, M via (Coughlin et al., 2008; Yang and Liu, 2003).

\begin{equation}
\frac{dm}{dt} = \frac{Mq}{3P(1-q^2)} \frac{dP}{dt}
\label{dmdt1}
\end{equation}

\noindent
By employing M = 1.62$\pm$0.09 M$_{\odot}$ and q = 0.238; the values obtained in section \ref{modeling}, mass transfer rate of \astrobj{FO Hya} between the components turns out to be 1.67 $\times$ 10$^{-8}$ M$_{\odot}$ yr$^{-1}$. \\

The sinusoidal term of the $(O-C)$ oscillation in equation (\ref{qsine}) reveals a periodic change with semi amplitude, A, of 0.0041 days and a period, $P_{3}$ of 42.30 years. The amplitude of period oscillation, $\Delta P$ may now be determined using the following relation due to Rovithis-Livaniou et al. (2000).

\begin{eqnarray}
\Delta P=\sqrt{2[1-\cos(2\pi P/P_{3})]} \times A
\label{delp}
\end{eqnarray}

\noindent
The amplitude of the period oscillation is obtained to be $\Delta P$ = 0.78 $\times 10^{-6}$ day leading to $\Delta P$/ $P$ = 1.66 $\times 10^{-6}$.\\

\begin{table*}
\begin{center}
\caption{Times of light minima of \astrobj{FO Hya}.}
\label{table3}
\begin{tabular}{p{1.0in}p{0.5in}p{0.6in}p{0.5in}p{1.0in}p{0.5in}p{0.6in}p{0.5in}}
\hline 
HJD & Type$^1$ & Method$^2$ & Ref.$^3$ & HJD        & Type$^1$ & Method$^2$ & Ref.$^3$ \\
(2400000+ ) &  &      &       & (2400000+) &          &        &      \\ 
\hline
31216.17000 & p & vis & (1) & 53123.54854 & s & ccd & (7) \\
31224.21000 & p & vis & (1) & 53129.65254 & s & ccd & (7) \\
31231.23000 & p & vis & (1) & 53133.64468 & p & ccd & (7) \\
44618.23440 & p & pe  & (2) & 53393.78252 & p & ccd & (7) \\
44620.11110 & p & pe  & (2) & 53400.82116 & p & ccd & (7) \\
44632.09000 & s & pe  & (2) & 53417.72362 & p & ccd & (7) \\
44632.32320 & p & pe  & (2) & 53436.74512 & s & ccd & (7) \\
44634.20050 & p & pe  & (2) & 53499.42900 & p & V   & (8) \\
44638.19000 & s & pe  & (2) & 53500.60707 & s & ccd & (7) \\
44640.30510 & p & pe  & (2) & 53504.59680 & p & ccd & (7) \\
44642.18270 & p & pe  & (2) & 53556.48409 & s & ccd & (7) \\
44649.22710 & p & pe  & (2) & 53673.86840 & s & ccd & (7) \\
44656.26980 & p & pe  & (2) & 53720.82771 & s & ccd & (7) \\
44659.08900 & p & pe  & (2) & 53724.81814 & p & ccd & (7) \\
46137.24970 & p & pe  & (2) & 53792.67499 & s & ccd & (7) \\
46138.19050 & p & pe  & (2) & 53836.57843 & p & ccd & (7) \\
46144.07000 & s & pe  & (2) & 53867.56291 & p & ccd & (7) \\
46144.29840 & p & pe  & (2) & 53871.56083 & s & ccd & (7) \\
47970.18000 & s & pe  & (2) & 53903.48665 & s & ccd & (7) \\
47971.11000 & s & pe  & (2) & 54115.26260 & s & Ic  & (9) \\
47974.15940 & p & pe  & (2) & 54157.75581 & p & ccd & (7) \\
48300.27000 & s & pe  & (2) & 54464.84950 & p & ccd & (7) \\
48301.21000 & s & pe  & (2) & 54505.69046 & p & ccd & (7) \\
48302.14000 & s & pe  & (2) & 54932.28417 & s & ccd & (10) \\
48303.32030 & p & pe  & (2) & 54933.22186 & s & ccd & (10) \\
50189.00300 & p & vis & (3) & 54934.16650 & s & ccd & (10) \\
50189.00300 & p & V   & (4) & 55254.16519 & p & ccd & (10) \\
51286.87600 & p & ccd & (5) & 55254.39782 & s & ccd & (10) \\
51905.28800 & p & V   & (5) & 55255.33811 & s & ccd & (10) \\
52347.14330 & p & ccd & (6) & 55257.22290 & s & ccd & (10) \\
52960.85457 & p & ccd & (7) & 55258.39126 & p & ccd & (10) \\
53079.64647 & p & ccd & (7) & 55258.16047 & s & ccd & (10) \\
53101.71414 & p & ccd & (7) & 55259.32907 & p & ccd & (10) \\
53109.69754 & p & ccd & (7) &             &   &     &    \\
\hline 
\end{tabular} 
\end{center}
\tiny $^1$Type: p = primary minimum time, s = secondary minimum time. \\
\tiny $^2$Method: vis = visual observation, pe = photoelectric data, V = Johnson V filter , ccd = charge-coupled device, Ic = Cousins I filter. \\       
\tiny $^3$References: (1) Zessewitsch, V. P. (1954); (2) Candy and Candy (1997); (3) Nagai, Kazuhiro; (4) Nagai, Kazuo; (5) Paschke, Anton; (6) Kiyota, Seichiiro; (7) ASAS; Pojmanski (2002) (8) Sobotka, Petr (2007); (9) Nakajima, Kazuhir; (10) Present Study.
\end{table*}

The cyclic variation may be interpreted as resulting either from the magnetic activity of one or both components (Applegate, 1992), or from the light-time effect (LITE) through the presence of a tertiary companion (Irwin, 1952, 1959). The required variation of the quadruple momentum $\Delta Q$ so as to reproduce this cyclic change, may be calculated using the following equation due to Lanza and Rodon\`{o} (2002).

\begin{eqnarray}
\frac{{\Delta P}}{P}=-9\frac{\Delta Q}{Ma^{2}}
\label{deq}
\end{eqnarray}

Using equation (\ref{a1a2}) we get $\Delta Q$ as 2.08 $\times 10^{49}$ and 0.49 $\times 10^{49}$ g $\textrm{cm}^{2}$ for primary and secondary components, respectively. The $\Delta Q$ values are smaller than its typical value of $10^{51}-10^{52}$ g $\textrm{cm}^{2}$ for close binaries (Lanza and Rodon\`{o}, 1999), suggesting, that the magnetic mechanism may not suffice significantly to explain the cyclical period variation. Hence it turns out that the cyclical period change is more plausibly be interpreted as due to the presence of a third body. \\

The projected orbital semi- major axis, $a_{12}\textrm{sin}i\prime $, of the orbit of the third body is related to the semi-amplitude of the O-C oscillation via $a_{12}\textrm{sin}i\prime = A \times c $ (Rovithis-Livaniou et al., 2000), where $i\prime$ is the inclination of the orbit of the third component and \textit{c} is the speed of light. We obtain $a_{12}\textrm{sin}i\prime$ to be 0.71 AU. Now, it is straight- forward to obtain the mass function of the third body, $f(M_3)$ using the following well-known equation.

\noindent Here \textit{a} represents the separation of the components. Let us recall that \astrobj{FO Hya} is a spectroscopic binary and hence its radial velocity curve analysis, which has been made earlier by Siwak et al. (2010), may also be used to determine 'a'.
The well known radial velocity amplitude parameters K$_{1}$ and K$_{2}$ are related to the respective a$_{1}$ and a$_{2}$ via
 
\begin{equation}
K_{1 (2)} = \frac{2\pi}{P}\frac{a_{1 (2)}\textrm{sin}i}{\sqrt{1-e^{2}}}
\label{a1a2}
\end{equation}

\noindent 
Making use of the values obtained by Siwak et al. (2010); K$_{1}$ = (62.48$\pm$0.97) km/s, K$_{2}$ = (253.20$\pm$4.50) km/s and of the period and inclination as obtained by us we get a$_{1}$ = (0.59$\pm$0.02) R$_{\odot}$, a$_{2}$ = (2.40$\pm$0.06) R$_{\odot}$ leading to a = (2.99$\pm$0.06) R$_{\odot}$ and a$_{1}$/a$_{2}$ = M$_{2}$/M$_{1}$ = q = 0.238. Now, using Kepler's third law it is straight forward to determine M and thereby to obtain M$_{1}$ = 1.31$\pm$0.07 M$_{\odot}$ and M$_{2}$ = 0.31$\pm$0.11 M$_{\odot}$. \\ 

\begin{equation}
f(M_{3}) = \frac{4\pi^{2}}{G{P}_{3}^{2}}\times(a_{12}\sin{i\prime})^{3} \\ 
\label{fm}
\end{equation}

\noindent We obtain the mass function of the third body to be $f(M_{3})$ = 0.0002 M$_{\odot}$. The mass function $f(M_{3})$ may also be expressed via  

\begin{equation}
f(M_{3}) = \frac{(M_{3}\sin{i\prime})^{3}} {(M_{1}+M_{2}+M_{3})^{2}}
\label{thmass}
\end{equation}

\noindent
where $M_{1}$, $M_{2}$, and $M_{3}$ represent the masses of the eclipsing pair and the third companion, respectively and $G$ is the gravitational constant. Now if one assumes the third body orbit to be coplanar to the orbit of the eclipsing pair (i.e., $i\prime = 78.^\circ05(\pm0.^\circ43$), the value of the lowest mass of the third body is computed to be 0.10 M$_{\odot}$.  

\section{Light curves from present observations} 

Fig. 5 shows the folded light curves of \astrobj{FO Hya} in B, V, R and I bands. A rounded bottom primary (phase = 0.0) and almost flat secondary (phase = 0.5) eclipses were seen which could be due to a total-eclipse configuration of the system. The primary eclipse is obviously deeper than the secondary one. The difference between the primary and secondary minima is found in the range 0.26 - 0.30 magnitudes in B, V, R, and I bands. Therefore, primary minimum is deeper and occurs when the hotter star passes behind the cooler one. On the other hand, shallower secondary minimum occurs when the cooler star passes behind the hotter one. The difference between primary maximum (phase = 0.25) and primary minimum is about 0.50, 0.48, 0.48 and 0.46 magnitudes in B, V, R and I bands, respectively. However, the difference between secondary maximum (phase = 0.75) and secondary minimum ranges from 0.23 to 0.24 mag in B, V, R, and I bands, respectively. Similar differences between maximum and minimum has also been reported earlier by Candy and Candy (1997) and Siwak et al. (2010).

In W UMa-type systems, as both the stars are very close to each other, so, there is a continuous light variation outside the eclipses. Moreover, the stars will experience gravitational distortion and heating effects. These effects shows the asymmetric light curves of W UMa-type systems. The observed light curves of \astrobj{FO Hya} are found to be asymmetric around the primary and secondary maxima, with the secondary maximum brighter than the primary maximum. The magnitude difference between phase at 0.75, which is brighter, and phase at 0.25 ranges from 0.039 to 0.041 mag in B, V, R, and I bands. This is typically explained by the presence of a spots on the components of the binary system. This indicates that light curves of \astrobj{FO Hya} shows the  O'Connell effect. The O'connel effect has also been noticed in many W UMa-type binaries e.g. \astrobj{FG Hya} (Qian and Yang, 2005), \astrobj{CU Tau} (Qian et al., 2005), \astrobj{AH Cnc} (Qian et al., 2006), \astrobj{AD Cnc} (Yang and Liu, 2002a), \astrobj{QX And} (Qian et al., 2007), \astrobj{CE Leo} (Yang and Liu, 2002b; Kang et al., 2004), \astrobj{BX Peg} (Lee et al., 2004). 

\section{Estimation of basic parameters of \astrobj{FO Hya}}

We determine the effective temperature of the primary component employing the results due to Wang (1994) on contact binaries. First of all, we compute the color of the system using their equation (17), period-color relation, given as :

\begin{equation}
(B-V)_0 = 0.077-1.003  log P
\label{BV}
\end{equation}

\noindent
where orbital period P is in days. This relation yields the intrinsic color of the star system to be (B-V)$_{0}$ = 0.406. Now we use this value of (B-V)$_{0}$ in their Eq. (14) i. e.  

\begin{equation}
(B-V)_0 = \frac{3.970 - log T_{eff}}{0.310}
\label{teff}
\end{equation}

\noindent
and get T$_{eff}$ of the primary component to be 6982 K. On the other hand, the infrared color index (J-K), using 2MASS catalogue (Cutri et al., 2003) 
turns out to be 0.235. This value when used in the color-temperature calibration of Cox (2000) yields a temperature of 7000 K and a spectral type F0V for the primary component of the binary system. The interstellar reddening is negligible in near infrared (Pribulla et al., 2009). Therefore, $(J-K)$ color from 2MASS is reliable to derive the spectral type of FO Hya. \\

Rucinski and Duerbeck (1997) has obtained a calibration of absolute magnitude ($M_V$) of W UMa-type star systems in terms of their $(B-V)_0$ color and period as

\begin{equation}
M_v = -4.44 log P + 3.02(B-V)_0 + 0.12
\label{vbvp}
\end{equation}

\noindent
We obtain $M_V$ of \astrobj{FO Hya} to be 2.80 mag. Taking the visual magnitude (V) of \astrobj{FO Hya} as 11.05 and assuming the negligible reddening, the distance modulus of \astrobj{FO Hya} is calculated to be 8.25 mag leading to a distance of 446.68 pc for the \astrobj{FO Hya}.

\begin{figure}
\begin{center}
{\includegraphics[height=8.0cm,angle=-360]{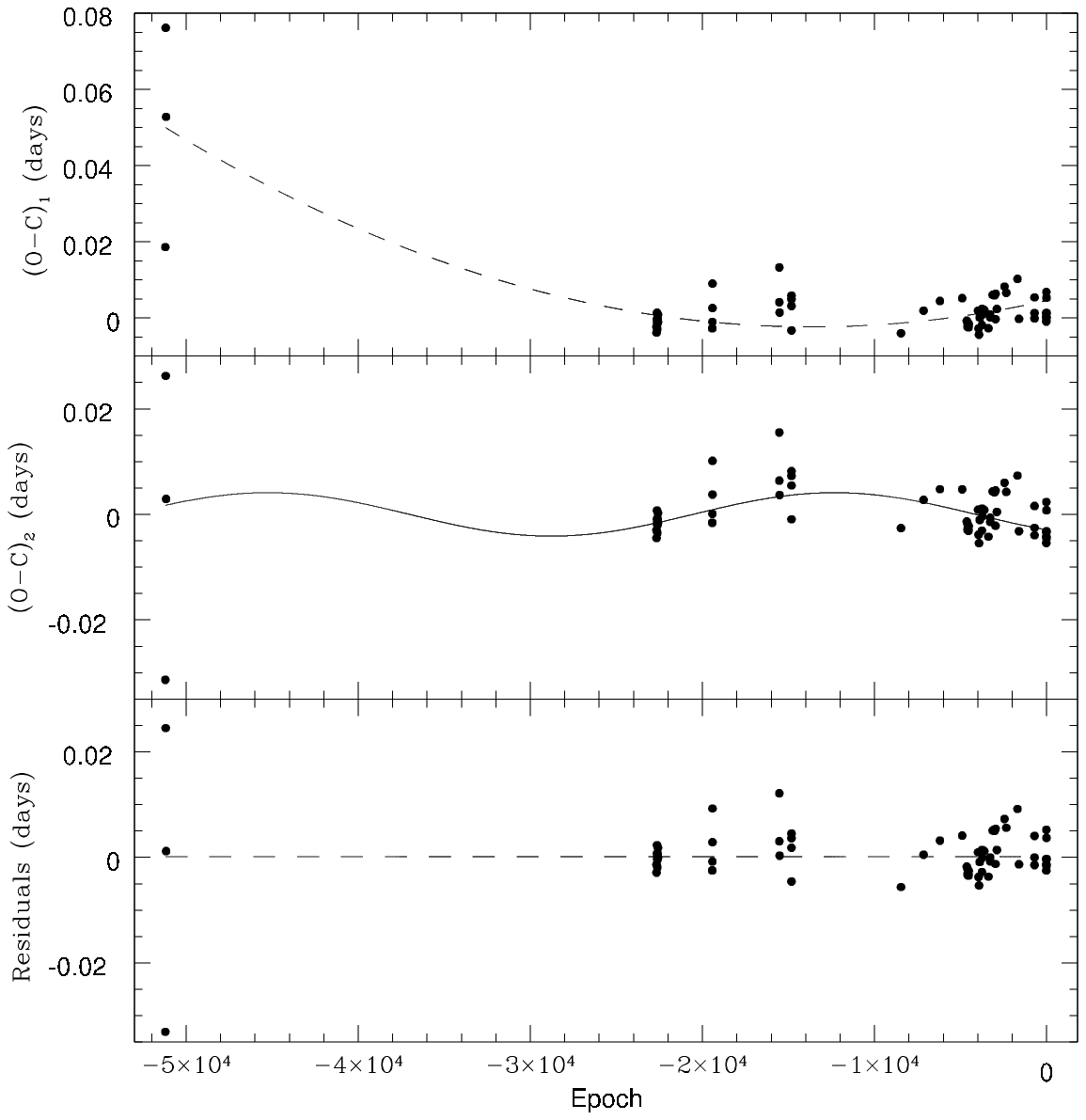}}
\caption {The top panel: (O-C)$_{1}$ diagram computed using equation (1). The general trend of the (O-C)$_{1}$ curve reveals a long-term period increase (dashed line). The middle panel: the (O-C)$_{2}$ curve from the parabolic fit and its description by a sinusoidal equation (solid line). The bottom panel: residuals from equation (2).}
\label{smfit}
\end{center}
\end{figure}
\section{Light Curve Modeling and Photometric Solutions}
\label{modeling}
We have modeled B, V, R and I band light curves using the WD code (Wilson and Devinney, 1971) implemented in PHOEBE\footnote{http://phoebe.fiz.uni-lj.si/} (Pr{\v s}a and Zwitter, 2005). It is a modified package of the widely used WD program for deriving the geometrical and physical parameters of the eclipsing binary stars. In the WD code, some of the parameters need to be fixed and the convergence of the solutions are obtained using the methods of multiple-subsets (Wilson and Biermann, 1976). Since most of the W UMa binaries are the contact binaries, therefore, mode-3 of WD code is the most suitable for the light curve analysis. However, we start with mode-2 (for detached binary) and achieve the convergence of the solution and thereafter go to mode-3 for the final convergence. In the light curve analysis using the WD code, star 1 is the one eclipsed at primary minimum and star 2, at the secondary minimum. This notation has been used throughout the modeling. 

\begin{table*}
\begin{center}
\caption{Photometric solutions obtained from the light curve modeling of the \astrobj{FO Hya}.}
\label{table4} 
\begin{tabular}{p{2.5in} p{1.8in} p{1.8in}}
\hline
Parameters & Symbol & Photometric elements \\
\hline 
Mass Ratio (M$_{2}$/M$_{1}$) & q & 0.238$\pm$0.006  \\ 
Orbital inclination [$^\circ$] & i & 78.05$\pm$0.43 \\ 
Relative luminosity of star 1 (B) & [L$_{1}$/(L$_{1}$ + L$_{2}$)] & 0.922 \\
                              (V) & [L$_{1}$/(L$_{1}$ + L$_{2}$)] & 0.923 \\ 
                              (R) & [L$_{1}$/(L$_{1}$ + L$_{2}$)] & 0.921 \\ 
                              (I) & [L$_{1}$/(L$_{1}$ + L$_{2}$)] & 0.908 \\ 
Monochromatic ld coefficients (B) & (x$_{1B}$ = x$_{2B}$) & 0.192 \\
                                  & (y$_{1B}$ = y$_{2B}$) & 0.690 \\ 
                              (V) & (x$_{1V}$ = x$_{2V}$) & 0.063 \\ 
                                  & (y$_{1V}$ = y$_{2V}$) & 0.724 \\ 
                              (R) & (x$_{1R}$ = x$_{2R}$) & 0.044 \\ 
                                  & (y$_{1R}$ = y$_{2R}$) & 0.736 \\ 
                              (I) & (x$_{1I}$ = x$_{2I}$) & 0.105 \\ 
                                  & (y$_{1I}$ = y$_{2I}$) & 0.698 \\ 
Bolometric ld coefficients & (x$_{1bol}$ = x$_{2bol}$) & 0.085 \\
                           & (y$_{1bol}$ = y$_{2bol}$) & 0.638 \\
Third light (B) & l$_{3B}$ & 0.05901$\pm$0.00003 \\
(V) & l$_{3V}$ & 0.05169$\pm$0.00002 \\
(R) & l$_{3R}$ & 0.03367$\pm$0.00009 \\
(I) & l$_{3I}$ & 0.03454$\pm$0.00003 \\
Surface Potentials & ($\Omega$$_{1}$ = $\Omega$$_{2}$) & 2.2211$\pm$0.0001 \\ 
Inner contact surface & ($\Omega$$_{in}$) & 2.3247 \\
Outer contact surface & ($\Omega$$_{out}$) & 2.1742 \\               
Fill-out factor & f & 0.68 \\
Effective temperatures & T$_{1}$ (K) & 7000 \\
                       & T$_{2}$ (K) & 5213.1$\pm$0.2 \\ 
Surface  albedos (bolometric) & A$_{1}$ = A$_{2}$ & 0.50 \\ 
Gravity darkening Coefficients & g1 = g2 & 0.32 \\ 
Synchronicity parameters & F$_{1}$=F$_{2}$ & 1.0 \\ 
Relative Radii of components \\
(pole) & r$_{1pole}$ & 0.498$\pm$0.001 \\
         & r$_{2pole}$ & 0.272$\pm$0.002 \\ 
(side) & r$_{1side}$ & 0.547$\pm$0.002 \\
         & r$_{2side}$ & 0.288$\pm$0.002 \\
(back) & r$_{1back}$ & 0.579$\pm$0.002 \\
         & r$_{2back}$ & 0.356$\pm$0.006 \\
Spot parameters \\
Spot latitude & [$\phi$(rad)] & 1.571 \\        
Spot longitude & [$\theta$(rad)] & 1.289 \\    
Spot radius & [r$_{s}$(rad)] & 1.078 \\     
Spot temperature factor & (T$_{s}$ /T$_{*}$) & 0.890 \\
Absolute bolometric magnitudes & M$_{bol1}$ & 2.77 \\
                               & M$_{bol2}$ & 3.03 \\
Effective gravity of the components & log g$_{1}$ (cgs) & 4.14$\pm$0.07 \\
                                    & log g$_{2}$ (cgs) & 4.01$\pm$1.25 \\ 
Semi-major axis & [a$_{orb}$(R$_{sun}$)] & 2.99$\pm$0.06 \\ 
\hline
\end{tabular}
\end{center}
\end{table*}

The temperature of the star 1, T$_{1} = 7000 $K, the bolometric albedos A$_{1}$ = A$_{2}$ = 0.5 for convective envelopes (Ruci{\'n}ski, 1969) and gravity darkening coefficients, g$_{1}$ = g$_{2}$ = 0.32 for convective envelopes (Lucy, 1967) are input parameters. These have been kept fixed to obtain exact solution. The bolometric (x$_{1bol}$, y$_{1bol}$, x$_{2bol}$, y$_{2bol}$) and monochromatic (x$_{1}$, x$_{2}$, y$_{1}$, y$_{2}$) limb darkening coefficients of the components are interpolated using square root law from van Hamme (1993) tables. The adjustable parameters have been the temperature of the star 2 (T$_{2}$), orbital inclination (i), the surface potentials of both components, $\Omega$$_{1}$ and $\Omega$$_{2}$; $\Omega$$_{1}$ = $\Omega$$_{2}$ (for contact binaries), and monochromatic luminosity of star 1, L$_{1}$. During the photometric solution, a third object, with light l$_{3}$, has also been taken as an adjustable parameter in order to get better fit of the light curves. The third light l$_{3}$ is in the unit of total light as implemented in PHOEBE. However, determination of third light from the photometric light curve modeling sometimes affects the orbital inclination and amplitude of the light curve. But we can not ignore the third light l$_{3}$ for our present light curve modeling, because from our O-C curve analysis (see section~\ref {period}), we have found the evidence of a third light l$_{3}$.

In order to derive reliable physical parameters of binary stars from photometric light curve modeling one requires the spectroscopically determined mass ratio (Deb and Singh, 2011). The spectroscopic mass ratio q$_{spec}$ = 0.238 (Siwak et al., 2010) of \astrobj{FO Hya} has been adopted during the modeling of light curves. We find that the theoretical light curves do not fit very well the observed ones; especially around the maxima. Therefore, the spot models are adopted to fit the observed light curves. We consider the possible spot models of hot or cool spots on the primary and secondary component, respectively. We then carry forth by iteration, allowing the spot latitude (ranging from 0$^\circ$ at the north pole to 180$^\circ$ at the south pole), the longitude (ranging from 0$^\circ$ to 360$^\circ$, with 0$^\circ$ at the inner Lagrangian point, 180$^\circ$ at the back end, and increasing in the direction of rotation), the angular radius (where 90$^\circ$ covers exactly half the star), and the temperature factor (the ratio of the spot temperature to the underlying surface temperature), as well as T$_{2}$ to vary, until a satisfactory fit is found until any further corrections were less than the errors. At this point, having obtained both orbital and spot parameters, we solve it further for a final solution for folded un-binned light curve, allowing all previously mentioned parameters to vary, once again until any further corrections were less than the errors. It turns out that placing the spot on the secondary component is found to be better than the spot on primary. Therefore, we take this as our final (best fit) solution that corresponds to the minimum sum of the square of the residuals value among them. The best fit theoretical light curves (solid lines) along with the observed light curve are shown in Fig. 5. The corresponding geometric configurations of \astrobj{FO Hya} with a dark spot on the secondary component are plotted in Fig. 6 at phases ($\phi$) = 0.00, 0.25, 0.50 and 0.75, respectively. \\
\begin{figure}
\begin{center}
{\includegraphics[height=8.0cm,angle=-0]{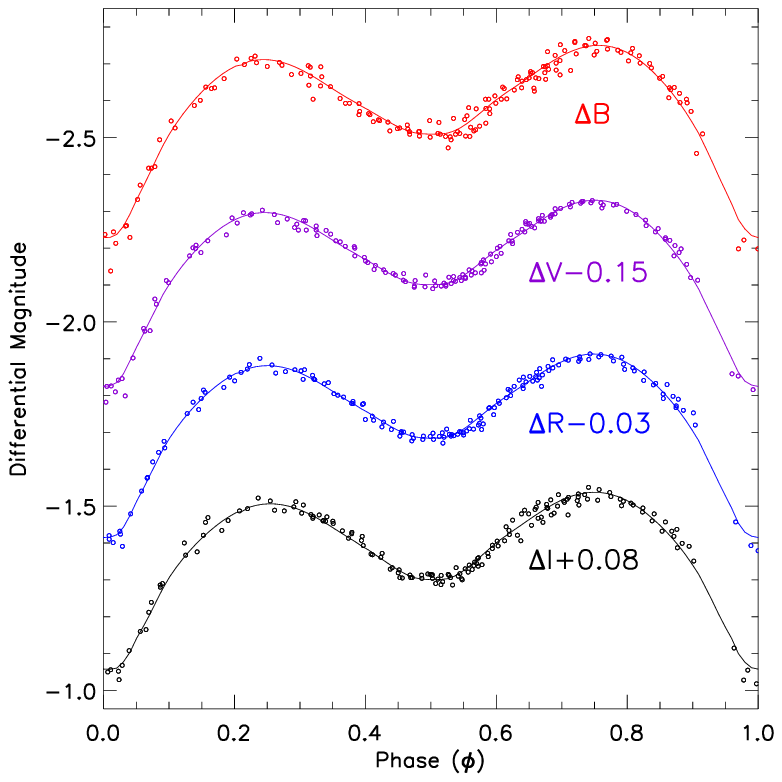}}
\caption {Phased light curves of the \astrobj{FO Hya} in B, V, R and I broad bands, respectively. Open circles denote the observational data points. The continuous line is the synthetic light curves computed from the WD light curve modeling technique considering the case as presence of spot on the secondary component.}
\label{fitt}
\end{center}
\end{figure}

The spot solutions included in the WD code can be easily fitted to the whole light curve but poses a serious problem regarding uniqueness of the solution (Maceroni and van't Veer, 1993) unless other means of investigation such as Doppler Imaging techniques are applied (Maceroni et al., 1994). The WD's differential correction ($dc$) minimization program yields the values of the fitting parameters as well as the formal statistical errors associated with each of them. We perform the Monte Carlo parameter scan (heuristic scan) around the best solution making use of the PHOEBE's scripter capability (Bonanos, 2009) with a view to explore the values, errors and stability of the solutions. We run the WD's ($dc$) minimization program 1000 times; updating each time the input parameter values for the next iteration. We obtain the final values for the parameters as the mean of the parameters resulted in various iterations. The errors on these parameters are derived by determining the standard deviations. Fig. 7 shows the histogram of the result obtained using the heuristic scan method for the orbital inclination, $i$ in the contact mode with PHOEBE for the \astrobj{FO Hya}. \\  

The light curve synthesis solutions from the present data indicate need of a third light. The presence of third light shows a tight correlation with the inclination angle and the mass ratio. It is remarked that correlation between mass ratio and third light still remains to be determined even for the best solutions available for the contact binaries showing total eclipses, where inner eclipse contacts can be well defined and the mass ratio can be determined solely from light curves (Mochancki and Doughty, 1972). During light curve modeling of the \astrobj{FO Hya} the mass ratio (q$_{spec}$) was fixed, therefore the third light highly correlated with the orbital inclination angle as shown in Fig. 7. In order to detect third light from photometric light curves with precision of $\approx$0.01 mag spectroscopic determination of the mass ratio is necessary (Pribulla and Rucinski, 2006). Precession of the orbital plane and an apsidal motion of the close binaries could be due to the third body orbiting a binary system. In eclipsing binaries, the change in orbital inclinations and consequently, change in the photometric amplitude of the eclipses are due their precession of the orbital plane. 

The best fit parameters are given in the Table \ref{table4}. The photometric elements such as T$_{2}$, i, $\Omega$$_{1}$, $\Omega$$_{2}$, relative luminosity and spot parameters obtained in the present investigation agree well with the previous solutions given by Candy and Candy (1997) and Siwak et al. (2010). The fractional stellar radii r$_{1,2}$ obtained from the light curve modeling are normalised to the semi-major axis of the binary system, i.e., r = R/a, where a and R are the semi-major axis and component radii, respectively. The component radius for each star (in solar radius units) may be determined using the relation R$_{1,2}$ = r$_{1,2}$a, where r is the geometrical mean of the r-pole, r-side and r-back radii for the \astrobj{FO Hya} contact system. From the modeling of the light curves, the temperature difference between the components of \astrobj{FO Hya} is found to be $\sim$ 1787 K. Because of this large temperature difference the light curves of \astrobj{FO Hya} show unequal depth. Such type of systems are known as B-type systems, which are in geometrical contact but not in thermal contact (e.g. Lucy and Wilson, 1979). These binaries show an EB ($\beta$ Lyrae)-type light curve. However, their orbital period falls in the range of classical W UMa systems (Ka{\l}u{\.z}ny, 1986). These type of systems are defined as B-type contact systems which show large differences in depths of eclipses. These systems are yet appear to be in contact and exhibit $\beta$ Lyrae-type light curves (Rucinski 1997). Some of them may be very close, semi-detached binaries mimicking contact systems and some may be in contact. \\ 

We determine the absolute physical parameters of the components, based on the results of the light curve solution. The luminosity of the star is calculated using the temperature and radius of each of the individual components. Table \ref{table5} lists the results regarding the physical parameters of the individual components of the \astrobj{FO Hya} system. \\
\begin{figure}
\begin{center}
{\includegraphics[height=3.0cm,angle=-360]{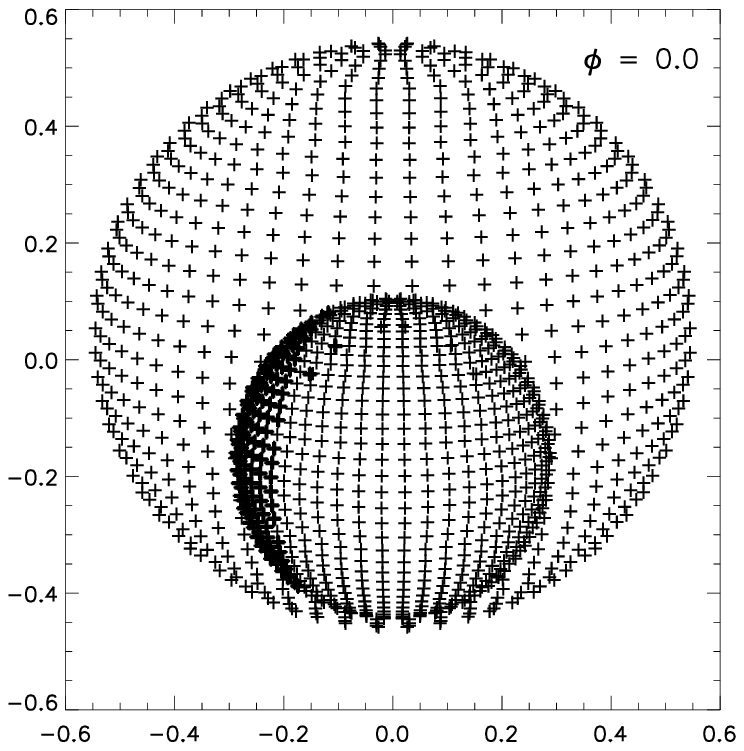}}    
{\includegraphics[height=3.0cm,angle=-360]{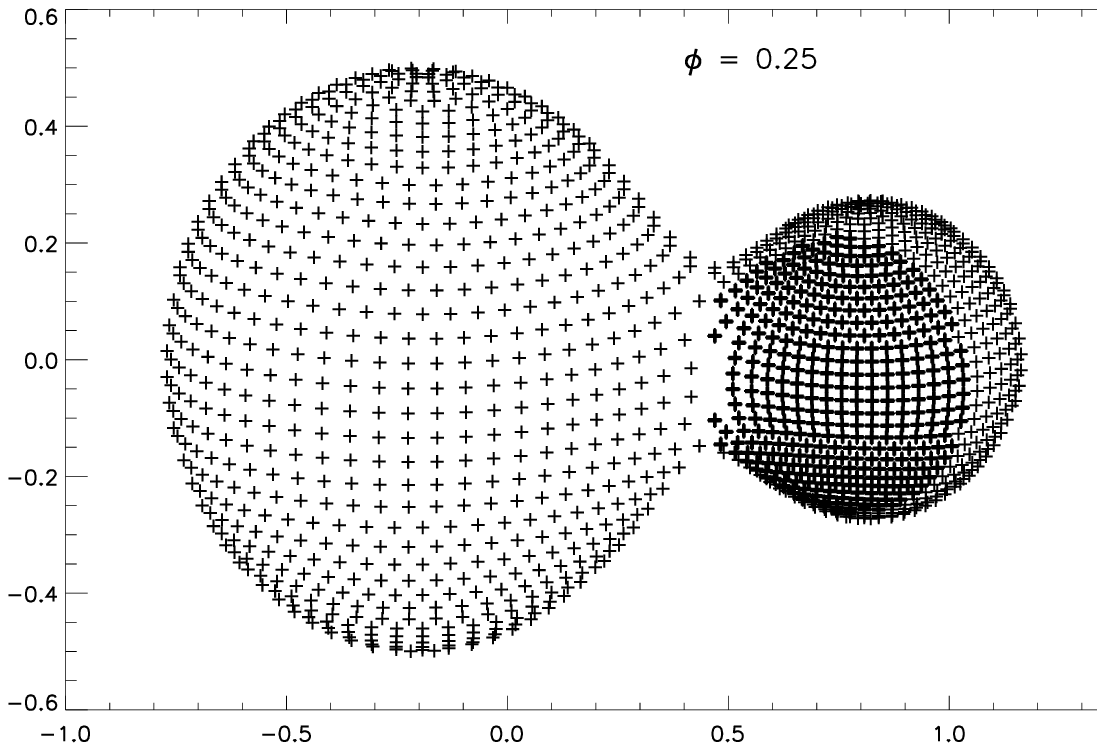}}   
{\includegraphics[height=3.0cm,angle=-360]{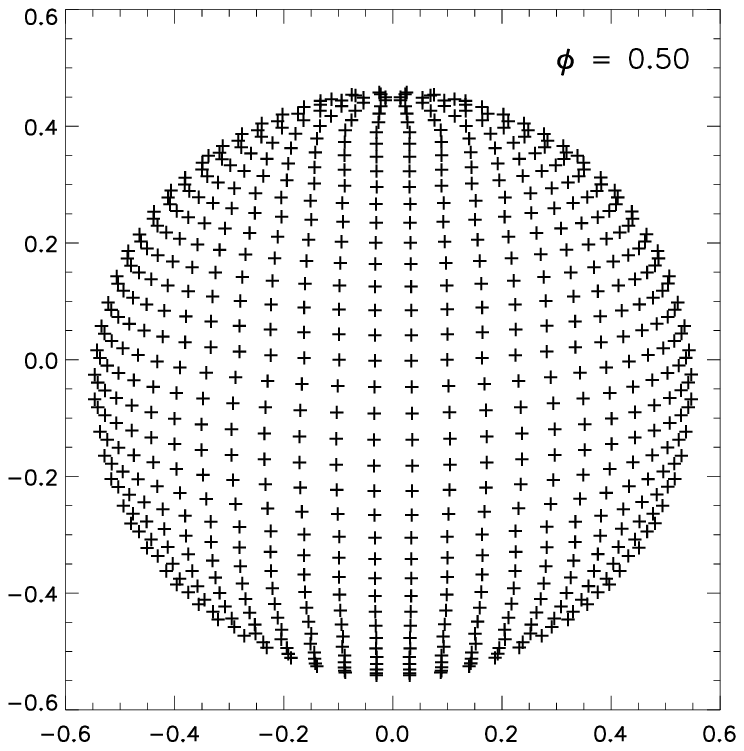}}
{\includegraphics[height=3.0cm,angle=-360]{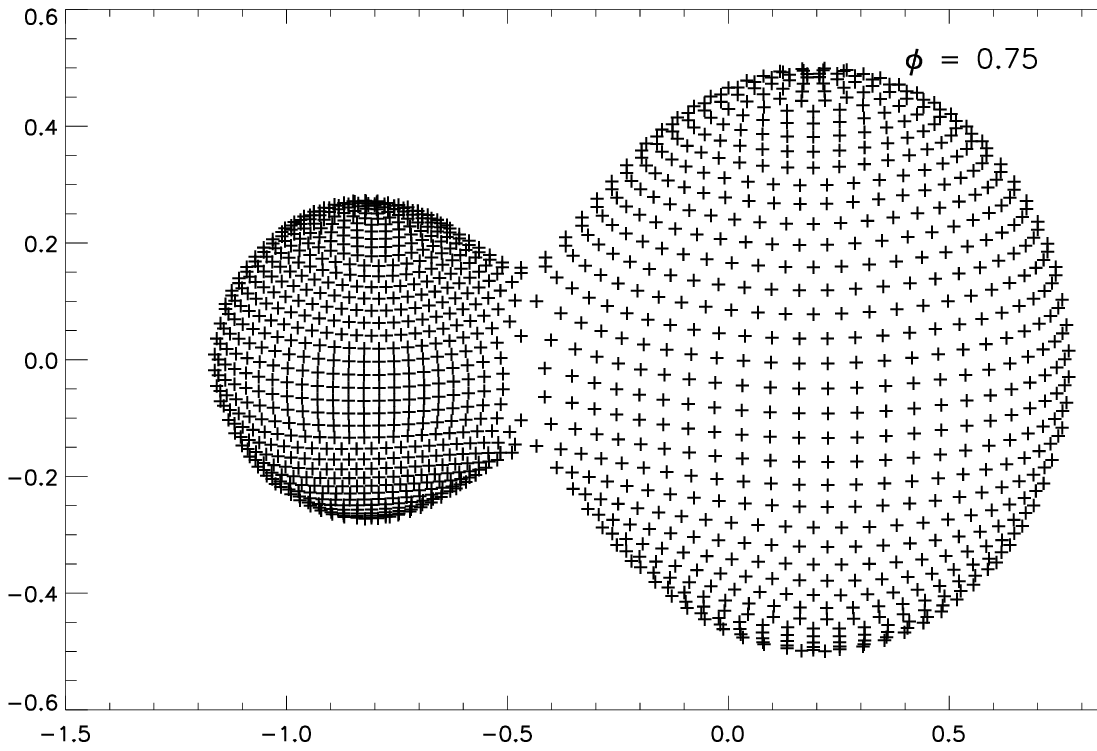}}
\caption {Geometric configurations of \astrobj{FO Hya} generated by PHOEBE at phases ($\phi$) = 0.00, 0.25, 0.50 and 0.75, respectively.}
\label{phase}
\end{center}
\end{figure}

The eclipsing binaries are classified into contact, semi-detached and detached binaries according to fill factor (f) as 0 $<$ f $<$ 1, f = 0 and f $<$ 0, respectively.
The fill-out factor (f) is given by
\begin{equation}
f = \frac{\Omega_{in} - \Omega_{1,2}}{\Omega_{in} - \Omega_{out}}
\label{fill}
\end{equation}

\noindent
where $\Omega$$_{in}$ and $\Omega$$_{out}$, and $\Omega_{1,2}$ are the inner and outer Lagrangian surface potentials, and surface potentials of the star 1 and 2, respectively. In case of contact binaries surface potentials of primary and secondary are equal to the surface potential ($\Omega$) of the common envelope for the binary system (i.e. $\Omega$$_{1}$ = $\Omega$$_{2}$ = $\Omega$). For semi-detached binaries with the primary and secondary components filling their Roche lobe, $\Omega$$_{1}$ = $\Omega$$_{in}$ and $\Omega$$_{2}$ = $\Omega$$_{in}$, respectively. For detached binaries $\Omega$$_{1,2}$ $>$ $\Omega$$_{in}$. The fill-out factor for the FO Hya is obtained to be 0.68 and is in the range of contact binaries. 

\begin{figure}[h]
\begin{center}
{\includegraphics[width=8.0cm,angle=-360]{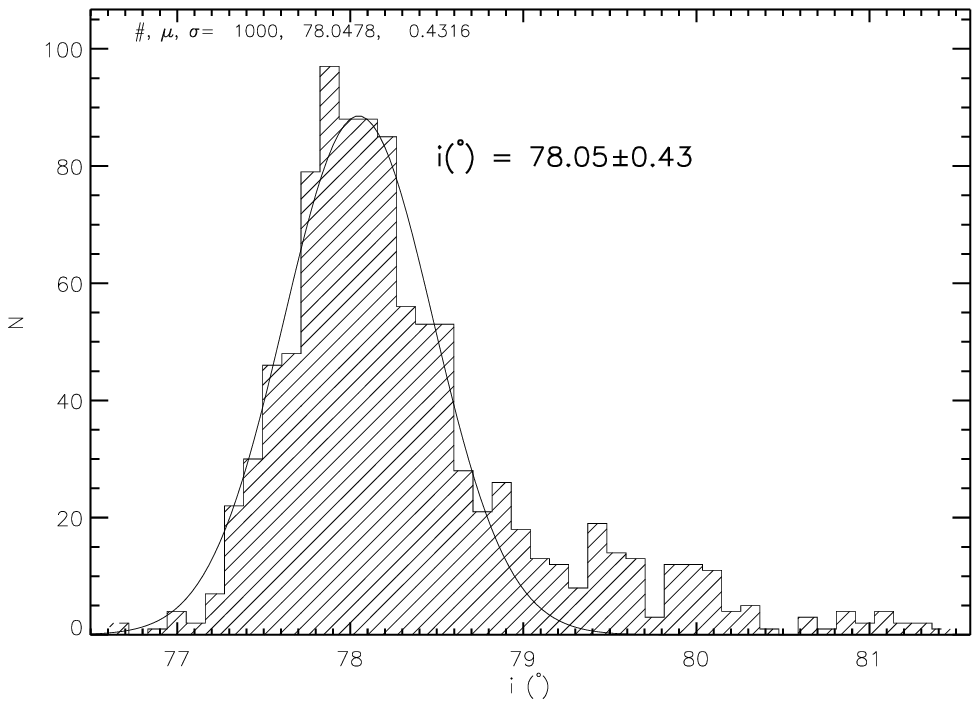}}
\caption {Histograms of the results obtained using the Monte Carlo parameter scan for the parameters fit of inclination in the contact mode with PHOEBE 
for the \astrobj{FO Hya}.}
\label{histi}
\end{center}
\end{figure}

\begin{table}[h]
\caption{Absolute parameters obtained from the light curve modeling.}
\label{table5}
\begin{tabular}{lcc}
\hline 
Parameters & Symbol & Photometric         \\ 
           &        &             elements\\ 
Mass \\
Primary & (M$_{1}$/M$_{\odot}$) & 1.31$\pm$0.07 \\
Secondary & (M$_{2}$/M$_{\odot}$) & 0.31$\pm$0.11 \\
Radius \\
Primary & (Mean Radius/R$_{\odot}$) & 1.62$\pm$0.03 \\
Secondary & (Mean Radius/R$_{\odot}$) & 0.91$\pm$0.02 \\
Luminosity \\
Primary & (L$_{1}$/L$_{\odot}$) & 5.65$\pm$0.21 \\
Secondary & (L$_{2}$/L$_{\odot}$) & 0.55$\pm$0.03 \\
\hline
\end{tabular}
\end{table}
The computed absolute parameters of \astrobj{FO Hya}, as reported in Table \ref{table5}, are used to estimate the evolutionary status of the system by means of the mass-radius diagram shown in Fig. 8. In this plot, the solid line represents the zero age main sequence (ZAMS) from Schaifers and Voigt, (1982). The position of components on the mass-radius diagram are obviously outside the zero age main sequence and they are located left on the zero age main sequence (ZAMS). It is evident from the Fig. 8 that the radius of the primary component approaches to the value for zero age main sequence (ZAMS) stars, i.e. basically a ZAMS star (Awadalla and Hanna, 2005; Liu and Yang, 2000b). Fig. 8 shows the mass-radius relation of the secondary is different from that of ZAMS stars (Yang and Liu, 2001), i.e. the secondary component is clearly far above the ZAMS than the location of primary, due to the fact that the mass transfer between the components will restructure the secondary and make it over-sized and over-luminous for its mass (Webbink, 2003; Li et al., 2008).

\begin{figure}[h]
\begin{center}
{\includegraphics[height=8.0cm,angle=-90]{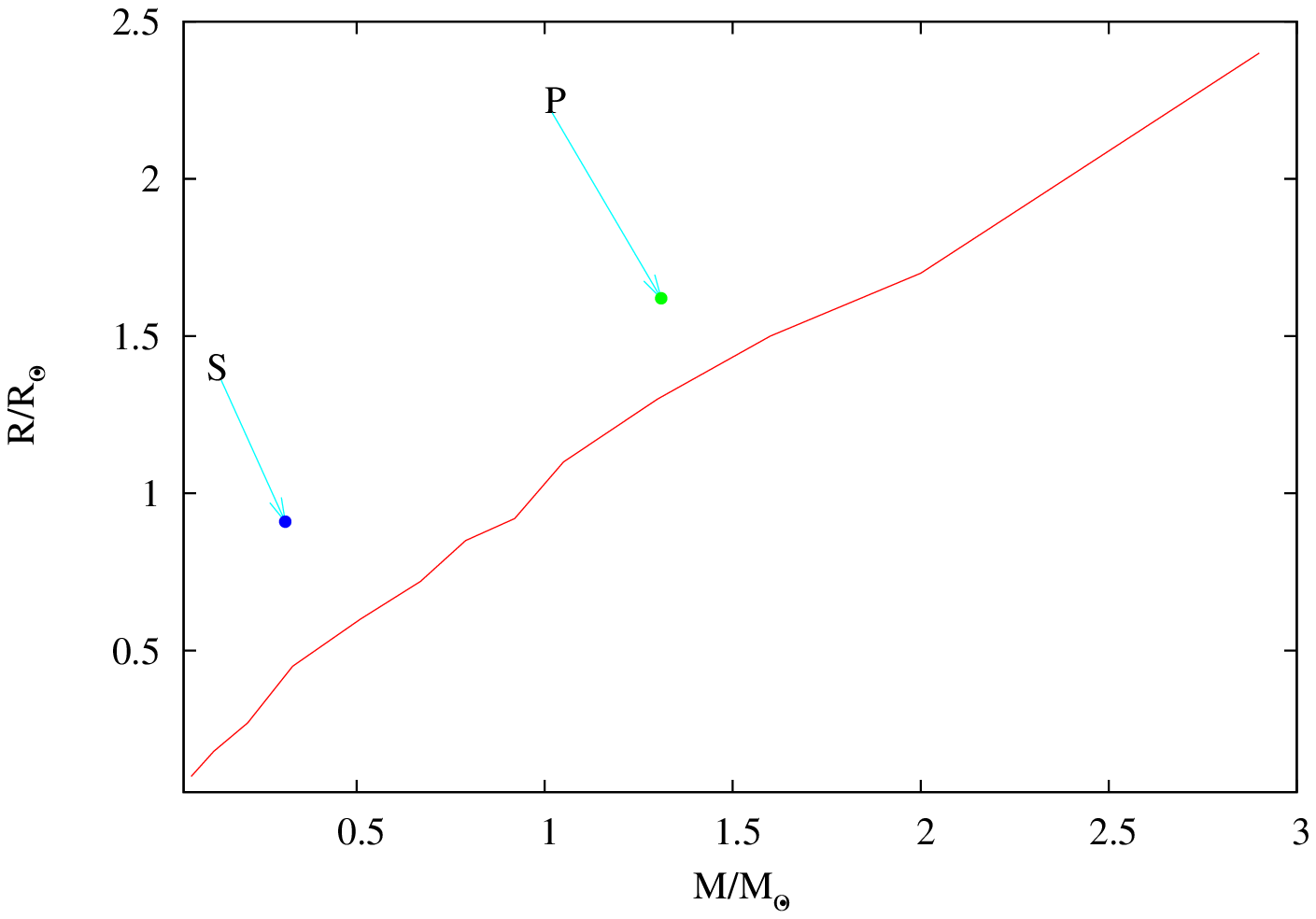}}
\caption {Location of the primary and secondary components of \astrobj{FO Hya} on a mass-radius diagram. The continuous line shows the zero age main sequence.}
\label{MR}
\end{center}
\end{figure}

\section{Sources of Polarization}

The values of polarization and polarization position angle of \astrobj{FO Hya} observed at three different epochs are given in Table \ref{table6}. Most of the polarimetric observations of \astrobj{FO Hya} have been taken during the phase from 0.13-0.26. However, the polarimetric observations in B band have been taken during the phase of 0.64. The degrees of polarization are found well with in $1 \sigma$ during these phases. Thus the observed polarization seems invariable. Since present observations are taken just before primary or secondary maxima so observed polarization may be invariable. The mean values of degree of polarization and polarization position angle were calculated to be 0.18$\pm$0.03, 0.15$\pm$0.03, 0.17$\pm$0.02 and 0.15$\pm$0.02 per cent and 61$\pm$6, 115$\pm$6, 89$\pm$3 and 98$\pm$5 degree in B, V, R and I bands, respectively. The degree of polarization in W UMa-type binaries were found upto 0.6\% (e.g. Oshchepkov, 1978; Pirrola, 1977). Oshchepkov (1978) found the variable polarization in W UMa type binaries and suggested that the photospheric scattering, the reflection effect and scattering in a gaseous envelope or stream could be the possible sources of variable polarization. However, Pirrola (1977) did not observe significant variability in the polarization and suggested that the polarization due to gas streams may be a transient phenomenon. More phase locked observations are needed to explain the observed polarization in \astrobj{FO Hya}. \\

Manset and Bastien (2000, 2001) have studied the polarimetric variations produced in a binary star by considering Thomson scattering and concluded that the average polarization produced depends on the grain's composition and size and on the wavelength of observation and they find that the highest polarizations are produced by grains with sizes in the range $\sim$ 0.1-0.2 $\mu$m.
\begin{table}[ht]
\scriptsize
\caption{Observed BVRI polarization values for \astrobj{FO Hya}.}
\label{table6}
\begin{tabular}{lccccc}
\hline
Date of  & Filter & P) & $\theta $ & JD& Phase \\ 
observations & & (\%) & $(^\circ)$ & (2455000+) & \\ \hline
April 07, 2010 & B & 0.14$\pm$0.09 & 62$\pm$11 & 294.15901 & 0.1752 \\
               & V & 0.17$\pm$0.02 & 55$\pm$3 & 294.16441 & 0.1867 \\
               & R & 0.12$\pm$0.05 & 149$\pm$6 & 294.17191 & 0.2027 \\
               & I & 0.11$\pm$0.04 & 18$\pm$5 & 294.16705 & 0.1923 \\
April 23, 2010 & V & 0.12$\pm$0.08 & 175$\pm$11 & 310.16255 & 0.2575 \\
               & R & 0.21$\pm$0.01 & 28$\pm$1 & 310.12981 & 0.1878 \\
               & I & 0.18$\pm$0.06 & 178$\pm$7 & 310.10323 & 0.1312 \\
May 20, 2010   & B & 0.22$\pm$0.01 & 60$\pm$1 & 337.10598 & 0.6382 \\
\hline
\end{tabular} 
\end{table}

\section{Summary}

We have carried out a detailed photometric analysis of the eclipsing binary \astrobj{FO Hya} using the data obtained from B, V, R and I band CCD photometric observations. Making use of the data available in the literature our analysis is based on the data for 65 years. Our observations yield the period to be 0.469556$\pm$0.000003 day, which is similar to that obtained previously. On the basis of the long term data, we found that the period of \astrobj{FO Hya} increases by 5.77 $\times 10^{-8}$ days yr$^{-1} $, probably caused by the mass transfer between the components at the rate of 1.67 $\times 10^{-8}$ M$_{\odot}$ yr$^{-1} $. In the O-C curve analysis, it is also found that the period of \astrobj{FO Hya} has varied in a sinusoidal way, superimposed on the long-term, upward parabolic variation. The period of the sinusoidal variation is about 42.30 years with an amplitude of 0.004 days. Further, the mass function and the mass of the third body are found to be 0.0002 M$_{\odot}$ and 0.10 M$_{\odot}$, respectively. The distance of \astrobj{FO Hya} is determined to be 446.68 pc. In the WD code analysis, we employed the mass ratio of 0.238$\pm$0.006 and found that the secondary component consist of dark spot on their surface. We have inferred the mass, radius and luminosity of primary and secondary components to be 1.31$\pm$0.07 M$_{\odot}$ and 0.31$\pm$0.11 M$_{\odot}$, 1.62$\pm$0.03 R$_{\odot}$ and 0.91$\pm$0.02 R$_{\odot}$, and 5.65$\pm$0.21 L$_{\odot}$ and 0.55$\pm$0.03 L$_{\odot}$, respectively. We have also derived filling factor to be 0.68. In this analysis we have taken up, presumably for the first time, the polarimetric studies of \astrobj{FO Hya}, and obtain the mean values of degree of polarization and polarization position angle to be 0.18$\pm$0.03, 0.15$\pm$0.03, 0.17$\pm$0.02 and 0.15$\pm$0.02 per cent and 61$\pm$6, 115$\pm$6, 89$\pm$3 and 98$\pm$5 degree in B, V, R and I bands, respectively. The level of optical polarization \astrobj{FO Hya} could be due to the presence of supplementary sources e.g. binarity and circumstellar material (gas streams). 
  
\section{Acknowledgments}

VP and DCS acknowledge the financial support received from UGC, Govt. of India under the fellowship (No. F. 14­2(SC)/2010 (SA­ΙΙΙ)). Authors acknowledge helpful discussions with Dr. Sukanta Deb. The authors would like to thank ARIES for making the telescope time available on the 104­cm Sampurnanand telescope and this research has made use of the data obtained from the 104­cm ST. VP also acknowledges to ARIES for library and computational facilities. The authors thank the staff at ARIES for their active support during the course of observations. The use of the SIMBAD, ADS, ESO DSS databases are gratefully acknowledged. The authors thank to the anonymous referee for insightful comments and suggestions.


\begin{thebibliography}{77}
\expandafter\ifx\csname natexlab\endcsname\relax\def\natexlab#1{#1}\fi
\expandafter\ifx\csname url\endcsname\relax
  \def\url#1{\texttt{#1}}\fi
\expandafter\ifx\csname urlprefix\endcsname\relax\def\urlprefix{URL }\fi

\bibitem[{{Akalin} and {Derman}(1997)}]{aka1997} 
{Akalin}, A., {Derman}, E., 1997. A$\&$AS 125, 407.

\bibitem[{{Applegate}(1992)}]{app1992} 
{Applegate}, J.~H., 1992. ApJ 385, 621.

\bibitem[{{Awadalla}(1994)}]{awa1994} 
{Awadalla}, N.~S., 1994. A$\&$A 289, 137.

\bibitem[{{Awadalla} and {Hanna}(2005)}]{awa2005} 
{Awadalla}, N.~S., {Hanna}, M.~A., 2005. Journal of Korean Astronomical Society 38, 43. 

\bibitem[{{Binnendijk}(1960)}]{bin1960} 
{Binnendijk}, L., 1960. AJ 65, 358.

\bibitem[{{Bonanos}(2009)}]{bon2009} 
{Bonanos}, A.~Z., 2009. ApJ 691, 407.

\bibitem[{{Candy} and {Candy}(1997)}]{can1997} 
{Candy}, M.~P., {Candy}, B.~N., 1997. MNRAS 286, 229.

\bibitem[{{Coughlin} et~al.(2008){Coughlin}, {Dale}, and {Williamon}}]{cou2008} 
{Coughlin}, J.~L., {Dale}, H.~A. III, {Williamon}, R.~M., 2008. AJ 136, 1089.

\bibitem[{{Cox}(2000)}]{cox2000} 
{Cox}, A.~N., 2000. {Allen's astrophysical quantities}.

\bibitem[{{Cutri} et~al.(2003){Cutri}, {Skrutskie}, {van Dyk}, {Beichman},
  {Carpenter}, {Chester}, {Cambresy}, {Evans}, {Fowler}, {Gizis}, {Howard},
  {Huchra}, {Jarrett}, {Kopan}, {Kirkpatrick}, {Light}, {Marsh}, {McCallon},
  {Schneider}, {Stiening}, {Sykes}, {Weinberg}, {Wheaton}, {Wheelock}, and
  {Zacarias}}]{cut2003}
{Cutri}, R.~M., {Skrutskie}, M.~F., {van Dyk}, S., {Beichman}, C.~A.,
  {Carpenter}, J.~M., {Chester}, T., {Cambresy}, L., {Evans}, T., {Fowler}, J.,
  {Gizis}, J., {Howard}, E., {Huchra}, J., {Jarrett}, T., {Kopan}, E.~L.,
  {Kirkpatrick}, J.~D., {Light}, R.~M., {Marsh}, K.~A., {McCallon}, H.,
  {Schneider}, S., {Stiening}, R., {Sykes}, M., {Weinberg}, M., {Wheaton},
  W.~A., {Wheelock}, S., {Zacarias}, N., 2003. {2MASS All Sky Catalog of point
  sources.}

\bibitem[{{Deb} and {Singh}(2011)}]{deb2011} 
{Deb}, S., {Singh}, H.~P., 2011. MNRAS 412, 1787.

\bibitem[{{Derman} and {Demircan}(1992)}]{der1992} 
{Derman}, E., {Demircan}, O., 1992. AJ 103, 1658.

\bibitem[{{Dryomova} and {Svechnikov}(2006)}]{dry2006} 
{Dryomova}, G.~N., {Svechnikov}, M.~A., 2006. Astrophysics 49, 358. 

\bibitem[{{Eggleton}(1996)}]{egg1996} 
{Eggleton}, P.~P., 1996. ASPC 90, 257.

\bibitem[{{Herczeg}(1993)}]{her1993} 
{Herczeg}, T.~J., 1993. PASP 105, 911.

\bibitem[{{Hilditch} and {King}(1988)}]{hil1988} 
{Hilditch}, R.~W., {King}, D.~J., 1988. MNRAS 231, 397.

\bibitem[{{Hoffmeister}(1936)}]{hof1936} 
{Hoffmeister}, C., 1936. AN 258, 39.

\bibitem[{{Irwin}(1952)}]{irw1952} 
{Irwin}, J.~B., 1952. ApJ 116, 211.

\bibitem[{{Irwin}(1959)}]{irw1959} 
{Irwin}, J.~B., 1959. AJ 64, 149.

\bibitem[{{Kang} et~al.(2004){Kang}, {Lee}, {Hong}, {Kim}, and {Guinan}}]{kan2004} 
{Kang}, Y.~W., {Lee}, H.~-W., {Hong}, K.~S., {Kim}, C.~-H., {Guinan}, E.~F., 2004. AJ 128, 846.

\bibitem[{{Kalimeris} et~al.(1994){Kalimeris}, {Rovithis-Livaniou}, {Rovithis}, {Oprescu}, {Dumitrescu}, and {Suran}}]{kal1994} 
{Kalimeris}, A., {Rovithis-Livaniou}, H., {Rovithis}, P., {Oprescu}, G., {Dumitrescu}, A., {Suran}, M.~D., 1994. A$\&$A 291, 765.

\bibitem[{{Kaszas} et~al.(1998){Kaszas}, {Vinko}, {Szatmary}, {Hegedus}, {Gal}, {Kiss}, and {Borkovits}}]{kas1998} 
{Kaszas}, G., {Vinko}, J., {Szatmary}, K., {Hegedus}, T., {Gal}, J., {Kiss}, L.~L., {Borkovits}, T., 1998. A$\&$A 331, 231.

\bibitem[{{Ka{\l}u{\.z}ny}(1986)}]{kau1986} 
{Ka{\l}u{\.z}ny}, J., 1986. AcA 36, 113.

\bibitem[{{Kim} et~al.(1997){Kim}, {Jeong}, {Demircan}, {Muyesseroglu}, and {Budding}}]{kim1997} 
{Kim}, C.~H., {Jeong}, J.~H., {Demircan}, O., {Muyesseroglu}, J., {Budding}, E., 1997. AJ 114, 2753.

\bibitem[{{Kopal}(1978)}]{kop1978} 
{Kopal}, Z., 1978. Ap$\&$SS 57, 439. 

\bibitem[{{Kwee} and {van Woerden}(1956)}]{kwe1956} 
{Kwee}, K.~K., {van Woerden}, H., 1956. Bull. Astron. Inst. Netherlands 12, 327. 

\bibitem[{{Lanza} and {Rodon\`{o}}(1999)}]{lan1999} 
{Lanza}, A.~F., {Rodon\`{o}}, M., 1999. A\&A 349, 887.

\bibitem[{{Lanza} and {Rodon\`{o}}(2002)}]{lan2002} 
{Lanza}, A.~F., {Rodon\`{o}}, M., 2002. AN 323, 424.

\bibitem[{{Lee} et~al.(2004){Lee}, {Kim}, {Han}, {Kim}, and {Koch}}]{lee2004} 
{Lee}, J.~W., {Kim}, C.~-H., {Han}, W., {Kim}, Ho-Il, {Koch}, R.~H., 2004. MNRAS 352, 1041.

\bibitem[{{Li} et~al.(2008){Li}, {Zhang}, {Han}, {Jiang}, and {Jiang}}]{liz2008} 
{Li}, L., {Zhang}, F., {Han}, Z., {Jiang}, D., {Jiang}, T., 2008. MNRAS 387, 97.

\bibitem[{{Liu} and {Yang}(2000a)}]{liu2000a} 
{Liu}, Q., {Yang}, Y., 2000a. A$\&$AS 142, 31.

\bibitem[{{Liu} and {Yang}(2000b)}]{liu2000b} 
{Liu}, Q., {Yang}, Y., 2000b. A$\&$A 361, 226.

\bibitem[{{Lucy} and {Wilson}(1979)}]{luc1979} 
{Lucy}, L.~B., {Wilson}, R.~E., 1979. ApJ 231, 502

\bibitem[{{Lucy}(1967)}]{luc1967} 
{Lucy}, L.~B., 1967. Zeitschrift fur Astrophysik 65, 89. 

\bibitem[{{Maceroni} and {van't Veer}(1993)}]{mac1993} 
{Maceroni}, C., {van't Veer}, F., 1993. A$\&$A 277, 515.

\bibitem[{{Maceroni} et~al.(1994){Maceroni}, {Vilhu}, {van't Veer}, and {van Hamme}}]{mac1994} 
{Maceroni}, C., {Vilhu}, O., {van't Veer}, F., {van Hamme}, W., 1994. A$\&$A 288, 529. 

\bibitem[{{Maddox} and {Bookmyer}(1981)}]{mad1981} 
{Maddox}, W.~C., {Bookmyer}, B.~B., 1981. PASP 93, 230.

\bibitem[{{Manset} and {Bastien}(2000)}]{man2000} 
{Manset}, N., {Bastien}, P., 2000. AJ 120, 413.

\bibitem[{{Manset} and {Bastien}(2001)}]{man2001} 
{Manset}, N., {Bastien}, P., 2001. AJ 122, 2692.

\bibitem[{{Medhi} et~al.(2007){Medhi}, {Maheswar}, {Brijesh}, {Pandey}, {Kumar}, and {Sagar}(2007)}]{med2007} 
{Medhi}, B.~J., {Maheswar}, G., {Brijesh}, K., {Pandey}, J.~C., {Kumar}, T.~S., {Sagar}, R., 2007. MNRAS 378, 881.

\bibitem[{{Medhi} et~al.(2008){Medhi}, {Maheswar}, {Pandey}, {Kumar} and {Sagar}(2008)}]{med2008} 
{Medhi}, B.~J., {Maheswar}, G., {Pandey}, J.~C., {Kumar}, T.~S., {Sagar}, R., 2008. MNRAS 388, 105.

\bibitem[{{Mochnacki} and {Doughty}(1972)}]{moc1972} 
{Mochnacki}, S.~W., {Doughty}, N.~A., 1972. MNRAS 156, 51.

\bibitem[{{O'Connell}(1951)}]{oco1951} 
{O'Connell}, D.~J.~K., 1951. Publications of the Riverview College Observatory 2, 85. 

\bibitem[{{Oshchepkov}(1978)}]{osh1978} 
{Oshchepkov}, V.~A., 1978. NInfo. 45, 500.

\bibitem[{{Pandey} et~al.(2009){Pandey}, {Medhi Biman}, {Sagar}, and {Pandey}}]{pan2009} 
{Pandey}, J.~C., {Medhi Biman}, J., {Sagar}, R., {Pandey}, A.~K., 2009. MNRAS 396, 1004.

\bibitem[{{Paschke} and {Brat}(2006)}]{pas2006} 
{Paschke}, A., {Brat}, L., 2006. OEJV 23, 13.

\bibitem[{{Pirrola}(1977)}]{pir1977} 
{Pirrola}, V., 1977. A$\&$A 56, 105. 

\bibitem[{{Pojmanski}(2002)}]{poj2002} 
{Pojmanski}, G., 2002. AcA 52, 397.

\bibitem[{{Pribulla} and {Rucinski}(2006)}]{pri2006}
{Pribulla}, T., {Rucinski}, S.~M., 2006. AJ 131, 2986.

\bibitem[{{Pribulla} et~al.(2009){Pribulla}, {Rucinski}, {Blake}, {Lu}, {Thomson}, {DeBond}, {Karmo}, {De Ridder}, {Ogłoza}, {Stachowski}, and {Siwak}}]{pri2009} 
{Pribulla}, T., {Rucinski}, S.~M., {Blake}, R.~M., {Lu}, W., {Thomson}, J.~R., {DeBond}, H., {Karmo}, T., {De Ridder}, A., {Ogłoza}, W., {Stachowski}, G., {Siwak}, M., 2009. AJ 137, 3655.

\bibitem[{{Pr{\v s}a} and {Zwitter}(2005)}]{prs2005} 
{Pr{\v s}a}, A., {Zwitter}, T., 2005. ApJ 628, 426.

\bibitem[{{Qian} and {Yang}(2005)}]{qia2005} 
{Qian}, S.~-B., {Yang}, Y.~-G., 2005. MNRAS 356, 765.

\bibitem[{{Qian} et~al.(2005){Qian}, {Yang}, {Soonthornthum}, {Zhu}, {He}, and {Yuan}}]{qi2005} 
{Qian}, S.~-B., {Yang}, Y.~-G., {Soonthornthum}, B., {Zhu}, L.~-Y., {He}, J.~-J., {Yuan}, J.~-Z., 2005. AJ 130, 224.

\bibitem[{{Qian} et~al.(2006){Qian}, {Liu}, {Soonthornthum}, {Zhu}, and {He}}]{qia2006} 
{Qian}, S.~-B., {Liu}, L., {Soonthornthum}, B., {Zhu}, L.~-Y., {He}, J.~-J., 2006. AJ 131, 3028.

\bibitem[{{Qian} et~al.(2007){Qian}, {Liu}, {Soonthornthum}, {Zhu}, and {He}}]{qia2007} 
{Qian}, S.~-B., {Liu}, L., {Soonthornthum}, B., {Zhu}, L.~-Y., {He}, J.-J., 2007. AJ 134, 1475.

\bibitem[{{Qian} and {Liu}(2000)}]{qia2000} 
{Qian}, S., {Liu}, Q., 2000. A$\&$A 355, 171.

\bibitem[{{Qian} and {Ma}(2001)}]{qia2001} 
{Qian}, S., {Ma}, Y., 2001. PASP 113, 754.

\bibitem[{{Rautela} et~al.(2004){Rautela}, {Joshi}, and {Pandey}}]{rau2004} 
{Rautela}, B.~S., {Joshi}, G.~C., {Pandey}, J.~C., 2004. BASI 32, 159.

\bibitem[{{Roberts} et~al.(1987){Roberts}, {Lehar}, and {Dreher}}]{rob1987} 
{Roberts}, D.~H., {Lehar}, J., {Dreher}, J.~W., 1987. AJ 93, 968.

\bibitem[{{Rovithis-Livaniou} et~al.(2000){Rovithis-Livaniou}, {Kranidiotis}, {Rovithis}, and {Athanassiades}}]{rov2000} 
{Rovithis-Livaniou}, H., {Kranidiotis}, A.~N., {Rovithis}, P., {Athanassiades}, G., 2000. A\&A 354, 904.

\bibitem[{{Rucinski}(1986)}]{ruc1986} 
{Rucinski}, S.~M., 1986. IAUS 118, 159.

\bibitem[{{Ruci{\'n}ski}(1969)}]{ruc1969} 
{Ruci{\'n}ski}, S.~M., 1969. AcA 19, 245. 

\bibitem[{{Rucinski}(1997)}]{ruc1997} 
{Rucinski}, S.~M., 1997. AJ 113, 407.

\bibitem[{{Rucinski} and {Duerbeck}(1997)}]{ruc1997b} 
{Rucinski}, S.~M., {Duerbeck}, H.~W., 1997. PASP 109, 1340.

\bibitem[{{Schaifers} and {Voigt}(1982)}]{sch1982} 
{Schaifers}, K., {Voigt}, H.~H., 1982. Journal of the British Astronomical Association 92, 154. 

\bibitem[{{Schimidt} et~al.(1992){Schimidt}, {Elston}, and {Lupie}}]{sch1992} 
{Schimidt}, G.~D., {Elston}, R., {Lupie}, O.~L., 1992. AJ 104, 1563.

\bibitem[{{Siwak} et~al.(2010){Siwak}, {Zola}, and {Koziel-Wierzbowska}}]{siw2010} 
{Siwak}, M., {Zola}, S., {Koziel-Wierzbowska}, D., 2010. AcA 60, 305.

\bibitem[{{Tsessevich}(1954)}]{tse1954} 
{Tsessevich}, V.~P., 1954. Izv. Astron. Obs. Odessa IV(II), 153.

\bibitem[{{van Hamme}(1993)}]{van1993} 
{van Hamme}, W., 1993. AJ 106, 2096.

\bibitem[{{Wang}(1994)}]{wan1994} 
{Wang}, J.~M., 1994. ApJ 434, 277.

\bibitem[{{Webbink}(2003)}]{web2003} 
{Webbink}, R.~F., 2003. ASPC 293, 76.

\bibitem[{{Wilson} and {Devinney}(1971)}]{wil1971} 
{Wilson}, R.~E., {Devinney}, E.~J., 1971. ApJ 166, 605.

\bibitem[{{Wilson} and {Biermann}(1976)}]{wil1976} 
{Wilson}, R.~E., {Biermann}, P., 1976. A$\&$A 48, 349. 

\bibitem[{{Yang} and {Liu}(2002a)}]{yan2002a} 
{Yang}, Y.~-L., {Liu}, Q.~-Y., 2002a. Chinese J. Astron. Astrophys. 2, 369. 

\bibitem[{{Yang} and {Liu}(2002b)}]{yan2002b} 
{Yang}, Y.~-L., {Liu}, Q.~-Y., 2002b. A$\&$A 387, 162.

\bibitem[{{Yang} and {Liu}(2001)}]{yan2001} 
{Yang}, Y., {Liu}, Q., 2001. AJ 122, 425.

\bibitem[{{Yang} and {Liu}(2003)}]{yan2003} 
{Yang}, Y., {Liu}, Q., 2003. PASP 115, 748. 

\bibitem[{{Zhai} et~al.(1984){Zhai}, {Leung}, and {Zhang}}]{zha1984} 
{Zhai}, D., {Leung}, K.~-C., {Zhang}, R., 1984. A$\&$AS 57, 487.

\end{thebibliography}
\end{document}